\documentstyle[12pt]{article}
\def\be{\begin{equation}}
\def\ee{\end{equation}}
\def\bea{\begin{eqnarray}}
\def\eea{\end{eqnarray}}
\def\a{\alpha}
\def\b{\beta}
\def\D{\Delta}

\author{Hans - J\"urgen Schmidt}

\title{The Newtonian limit of fourth-order gravity}

\date{}
\begin{document}
\maketitle

\centerline{Universit\"at Potsdam, Institut f\"ur Mathematik, Am
Neuen Palais 10} 
 \centerline{D-14469~Potsdam, Germany,  E-mail:
 hjschmi@rz.uni-potsdam.de}

\begin{abstract}
The weak-field slow-motion limit of fourth-order gravity will be discussed.
\end{abstract}

\bigskip

\noindent
Let us consider the gravitational theory defined by the Lagrangian
\be
 L_{\rm g}=  (8\pi G)^{-1} \left(
R/2 + (\a R_{ij}R^{ij} + \b R^2) l^2 \right) \, . 
\ee
$G$ is Newton's constant, $l$ a coupling length and $\a$
 and $\b$  numerical parameters. $R_{ij}$  and $R$  
are the Ricci tensor and its trace. Introducing the 
matter Lagrangian $L_{\rm m}$  and varying  $  L_{\rm g}
   +  L_{\rm m}$
 one obtains the field equation
\be
     E_{ij}  + \a H_{ij} + \b G_{ij}  = 8\pi G T_{ij}  \,   .         
\ee
For $\a = \b = 0$  this reduces to
 General Relativity Theory. The explicit expressions $H_{ij}$ and $ G_{ij}$
 can be found in  STELLE (1978).

\bigskip

In a well-defined sense, the weak-field slow-motion limit 
of Einstein's theory is just Newton's theory, 
cf.  DAUTCOURT (1964). In the following 
we consider the analogous problem for fourth 
order gravity eqs. (1), (2). For 
the special cases $\a = 0$  (PECHLANER,  
SEXL(1966),  POLIJEVKTOV-NIKOLADZE (1967)),  
$\a + 2\b = 0$  (HAVAS (1977),  JANKIEWICZ (1981)) and 
$\a  + 3\b  = 0$  (BORZESZKOWSKI, TREDER, 
YOURGRAU (1978))  this has already been done in the 
past. Cf. also ANANDAN (1983),  where torsion 
has been taken into account.

\bigskip

The slow-motion limit can be equivalently described 
as the limit $c \to \infty$, where $c$  is the  velocity of  light. 
In this sense we have to  take  the  limit 
$G \to  0$  while $G \cdot  c$  and $l$  remain constants. 
Then the energy-momentum tensor $T_{ij}$ 
 reduces to the rest mass density $\rho$:
\be
     T_{ij}=    \delta^0_i \delta^0_j \rho  \,    ,    
\ee
$x^0=     t $ being the time coordinate. The metric  can be written as
\be
     ds^2 =  (1 - 2\phi) dt^2
 - (1 + 2\psi) (dx^2 + dy^2 + dz^2)          \, .
\ee
Now eqs. (3) and (4) will be inserted into eq. (2). 
In our approach, products and time derivatives 
 of   $\phi$   and $\psi$ can be neglected,  i.e.,
$$
R=4 \D \psi -2 \D \phi \, ,    \qquad {\rm where} \qquad 
\D f =f_{,xx} + f_{,yy} + f_{,zz} \, .
$$
Further 
$R_{00} = - \D \phi$, $H_{00} = -2 \D R_{00} - \D R$
  and $G_{00} = -4 \D R$, where $l = 1$.

\bigskip

Then it holds: 
The validity of the $00$-component and of the trace of eq. (2),
\be
 R_{00} - R/2 + \a H_{00} + \b G_{00} = 8 \pi  G \rho    
\ee
and
\be
- R - 4(\a + 3 \b ) \D R = 8 \pi G \rho \,  ,
\ee
imply the validity of the full eq. (2).

\bigskip

Now, let us discuss eqs. (5) and (6) in more details: Eq. (5) reads
\be
- \D \phi - R/2 
+\a (2 \D \D \phi - \D R) - 4 \b \D R = 8 \pi G \rho \, .  
\ee
Subtracting one half of  eq. (6) yields
\be
- \D \phi   + 2 \a  \D \D \phi + (\a + 2 \b) \D R = 4 \pi G \rho \, .  
\ee
For $\a  + 2 \b  = 0$  one obtains
\be
- (1 -  2 \a  \D ) \D \phi   = 4 \pi G \rho 
\ee
and then   $\psi  = \phi$  is a solution of eqs. (5),  (6). 
For all other cases the equations for 
$\phi$  and $\psi $ do not decouple immediately, but, 
to get equations comparable with Poisson's equation 
we apply $\D$ to eq. (6)  and continue as follows.

\bigskip

     For $\a + 3\b  = 0$  one gets from eq. (8)
\be
- (1 -  2 \a  \D ) \D \phi   = 4 \pi G ( 1 + 2 \a \D /3 )  \rho   \, .
\ee
The $\D$-operator applied to the source term 
in eq. (10)  is only due to  the application  of 
$\D$ to the trace, 
the original equations (5), (6) contain only $\rho$ itself.

\bigskip

For $\a = 0$ one obtains similarily the equation
\be
- (1 + 12 \b  \D ) \D \phi   = 4 \pi G ( 1 + 16 \b \D )  \rho   \, .
\ee

\bigskip

For all other cases - just the cases not yet covered 
by the literature - the elimination of $\psi$ from  the 
system (5), (6) gives rise to a sixth-order equation
\be
- \left(1 + 4(\a + 3 \b)   \D \right)
 ( 1 - 2 \a \D)
  \D \phi   =
 4 \pi G \left( 1 + 2(3 \a + 8 \b) \D \right)  \rho   \, .
\ee

\bigskip

Fourth-order gravity is motivated by quantum-gravity 
considerations and therefore, its long-range behaviour 
should be the same as in Newton's theory. 
Therefore, the signs of the parameters $\a$, $\b$ should 
be chosen to guarantee an exponentially vanishing and 
not an oscillating behaviour of the fourth-order terms:
\be
\a \ge 0 \, , \qquad   \a + 3 \b \le 0 \, .
\ee
On the other hand, comparing parts of eq. (12)  with  
the Proca equation it makes sense to 
define the masses 
\be
m_2 = \left(2 \a  \right)^{-1/2} \quad {\rm and} \quad 
m_0 = \left(  -4(\a + 3\b )    \right)^{-1/2} \, .
\ee
Then (13) 
requires the masses of the spin 2 and spin 0 gravitons to be real.

\bigskip

Now, inserting a delta source $\rho  = m \delta $  into eq. (12)  one 
obtains for $\phi $ the same result as STELLE (1978), 
\be
\phi = m r^{-1} \left( 1 + \exp (-m_0 r)/3 - 4 \exp 
(-m_2 r)/3 \right) \, .
\ee
To obtain the metric completely one has also to calculate $\psi$.
 It reads
\be
\psi = m r^{-1} \left( 1 - \exp (-m_0 r)/3 - 2 \exp (-m_2 r)/3
 \right) \, .
\ee
For finite values $m_0$ and $m_2$  these are both 
bounded functions,  also for $r \to 0$. In the limits $\a \to  0$
 $(m_2 \to \infty)$    and
 $\a + 3\b \to 0$ $(m_0 \to \infty)$  the terms with   $m_0$ and $m_2$   
in eqs.  (15)  
and (16) simply vanish.  For these cases $\phi$  and $\psi $ become
 unbounded as $r \to 0$.

\bigskip

Inserting  (15), (16) into the metric  (4),  the behaviour of 
the geodesics shall be studied. First, for an estimation 
of the sign of the gravitational force we take a test 
particle at rest and look whether it starts falling towards 
the centre or not. The result is: for $m_0 \le 2 m_2$,  
gravitation is 
always attractive,  and for $m_0> 2m_2$  it is attractive 
for large but repelling for small distances.
 The intermediate case 
$m_0 = 2m_2$,  i.e., $3\a + 8\b =  0$, is
 already known to  be a special
one from eq. (12).

\bigskip

Next,  let us study the perihelion advance 
for distorted circle-like orbits. Besides the general 
relativistic perihelion advance (which vanishes 
in the Newtonian limit) we have an additional 
one of the following behaviour: 
For $r \to   0$  and $ r \to \infty $  it vanishes and for
$r \approx 1/m_0$ 
and $r \approx 1/m_2$  it has local maxima,  i.e., resonances.

\bigskip

Finally,  it should be mentioned that the 
gravitational field of an extended 
body can be obtained by integrating eqs. (15),  (16). 
For a spherically symmetric body 
the far field is also of  the 
type 
$$
 m r^{-1} \left( 1 + a \exp (-m_0 r) + b  \exp (-m_2 r) \right) \, ,
$$
 and the 
factors $a$ and $b$ carry 
information  about the mass distribution inside the body.

\section*{References}

\noindent
ANANDAN,  J.: 1983, 941 in: HU,  N., 
Proc. 3. Marcel Grossmann Meeting B, Amsterdam NHPC.

\noindent
BORZESZKOWSKI, H.,
 TREDER,  H.,  YOURGRAU,  W.: 1978,  Ann. Phys. Leipz. {\bf  35}, 
471.

\noindent
DAUTCOURT, G: 1964,  Acta Phys. Polon. {\bf  25}, 637.

\noindent
HAVAS, P.: 1977,   Gen. Rel. Grav. {\bf  8},  631.

\noindent
JANKIEWICZ,  Cz.: 1981,  Acta Phys. Polon. {\bf  13},  859.

\noindent
PECHLANER, E.  SEXL, R.: 1966,  Commun. Math. Phys. {\bf 2}, 165.

\noindent
POLIJEVKTOV-NIKOLADZE, N.: 1967,  J. exp. i  teor.
 Fiziki {\bf  52},  1360.

\noindent
STELLE, K.: 1978,  Gen. Rel. Grav. {\bf  9}, 353.

\bigskip

\medskip
\noindent
(Received 1986 January, 5)

\medskip

\noindent 
{\small {This is a  reprint from Astronomische Nachrichten, 
done with the kind permission of the copyright owner, only some obvious
 misprints have been cancelled;    
 Astron. Nachr. {\bf 307} (1986) Nr. 5, pages 339 - 340;  
  Author's address that time:  
Zentralinstitut f\"ur  Astrophysik der AdW der DDR, 
1502 Potsdam--Babelsberg, R.-Luxemburg-Str. 17a.}}

\end{document}